# Simplified configuration of Brillouin optical correlation-domain reflectometry

**Neisei Hayashi, Yosuke Mizuno,** *Member IEEE*,
**and Kentaro Nakamura,** *Member IEEE*

*Precision and Intelligence Laboratory, Tokyo Institute of Technology,*
*4259 Nagatsuta-cho, Midori-ku, Yokohama 226-8503, Japan*

*DOI: 10.1109/JPHOT.2014.xxxxxxx*
*xxxx-xxxx (c) 2014 IEEE. Translations and content mining are permitted for academic research only.*
*Personal use is also permitted, but republication/redistribution requires IEEE permission.*
*See http://www.ieee.org/publications_standards/publications/rights/index.html for more information.*

Manuscript received September xx, 2014; revised xxxxx xx, 2014; accepted xxxxx xx, 2014. Date of publication xxxxx xx, 2014; date of current version xxxxx xx, 2014. This work was partially supported by Grants-in-Aid for Young Scientists (A) (no. 25709032) and for Challenging Exploratory Research (no. 26630180) from the Japan Society for the Promotion of Science (JSPS) and by research grants from the Iwatani Naoji Foundation and the SCAT Foundation. N. Hayashi acknowledges a Grant-in-Aid for JSPS Fellows (no. 25007652). Corresponding author: N. Hayashi (e-mail: hayashi@sonic.pi.titech.ac.jp).

**Abstract:** We develop a simple and cost-efficient configuration of Brillouin optical correlation-domain reflectometry (BOCDR), the setup of which does not include an additional reference path used in standard BOCDR systems. The Fresnel-reflected light from an open end of a sensing fiber is used as reference light. The limitations of spatial resolution, measurement range, and their ratio are theoretically clarified, and then a distributed strain measurement with a < 100-mm spatial resolution and a 4.1-m measurement range is demonstrated with a high signal-to-noise ratio.

## 1. Introduction

Optical fiber sensing based on Brillouin scattering has emerged as a promising technique for monitoring conditions in diverse materials and structures, owing to its measurability of strain and/or temperature distribution along a fiber. To date, a variety of distributed sensing systems have been implemented, which can be basically categorized into five techniques: Brillouin optical time-domain reflectometry (BOTDR) [1-4], Brillouin optical time-domain analysis (BOTDA) [5-11], Brillouin optical frequency-domain analysis (BOFDA) [12-15], Brillouin optical correlation-domain analysis (BOCDA) [16-19], and Brillouin optical correlation-domain reflectometry (BOCDR) [20-26]. BOTDR and BOTDA have the measurement range as long as several tens of kilometers or longer, but the spatial resolution of basic time-domain techniques is inherently limited to ~1 m [27, 28], though several methods have recently been developed to enhance the resolution [2-4, 7-11]. BOFDA is free from the Brillouin linewidth broadening at high spatial resolution (~3 cm [14]), and BOCDA has an extremely high spatial resolution (~1.6 mm [17]). However, in standard BOFDA and BOCDA, two light beams (pump and probe) need to be injected into both ends of the sensing fiber, resulting in their complicated setup and/or high cost. Even their linear configurations [15, 19] do not ameliorate the situation. To resolve this problem, BOCDR was proposed in 2008.

BOCDR [20-26], which operates based on the correlation control of continuous waves (Brillouin Stokes light and reference light) [29], has a substantial one-end accessibility, a high spatial resolution (13 mm with a silica fiber [21] and 6 mm with a tellurite fiber [23]), and a high sampling rate. Although the spatial resolution and the measurement range are in the trade-off relation, temporal gating [24] and dual modulation [25] techniques have been developed to mitigate this drawback. Another practical advantage of BOCDR is a system simplicity and cost efficiency; the experimental setup of its basic configuration does not include relatively expensive devices, such as electro-optic modulators (optical-pulse generators, single-sideband modulators, etc) and vector network analyzers (cf. BOFDA). Further simplification of the BOCDR system will greatly enhance its utility.

In this work, we develop an even simpler experimental setup of BOCDR, named S-BOCDR, which exploits as reference light the Fresnel-reflected light from an open end of a sensing fiber and does not include an additional reference path used in standard implementations. After clarifying the theoretical limitations of spatial resolution, measurement range, and their ratio, we demonstrate a distributed strain measurement with a < 100-mm spatial resolution and a 4.1-m measurement range.

## 2. Principle and theory

When pump light is launched into an optical fiber, it interacts with acoustic phonons, generating backscattered Stokes light. This phenomenon is known as spontaneous Brillouin scattering [30], and the Stokes light spectrum, referred to as the Brillouin gain spectrum (BGS), takes the shape of a Lorentzian function. The central frequency of the BGS is known to shift downward relative to that of the pump light by the amount called the Brillouin frequency shift (BFS). When the pump wavelength is 1.55 μm, the BFS in silica single-mode fibers (SMFs) is approximately 10.8 GHz, which slightly varies depending on the fiber fabrication process. If a strain (or temperature change) is applied to a silica SMF, the BFS shifts toward a higher frequency with +580 MHz/% [31] (or +1.18 MHz/K [32]) at 1.32 μm, corresponding to +493 MHz/% (or +1.00 MHz/K) at 1.55 μm. Thus, by measuring the BFS distribution along a fiber, the information on applied strain or temperature distribution can be obtained.

BOCDR is one of the Brillouin-based distributed sensing techniques. Figure 1 depicts a schematic of the basic BOCDR system [20]. The Stokes light returned from a fiber under test (FUT) is mixed with the reference light, and their beat signal is converted into an electrical signal with a photo diode (PD) and is detected as a BGS with an electrical spectrum analyzer (ESA). By modulating the laser frequency with a sinusoidal waveform, correlation peaks are periodically synthesized along the FUT [29]. The FUT length is often set to be shorter than the correlation peak interval so that only one correlation peak is in the FUT. By controlling the optical path difference (e.g., by changing the reference path length), a correlation peak of any order can be located in the FUT. The position of the correlation peak can be scanned along the FUT by sweeping the frequency of the sinusoidal modulation, and thus, a distributed BGS measurement is feasible. The measurement range of the basic BOCDR, corresponding to the interval of the two neighboring correlation peaks, is expressed as [22]

$$d_{\text{BOCDR}} = \frac{c}{2 n f_m}, \quad (1)$$

where $c$ is the light velocity in vacuum, $n$ is the core refractive index, and $f_m$ is the modulation frequency of the laser frequency. In the meantime, the spatial resolution is given by [22]

$$\Delta z_{\text{BOCDR}} = \frac{c \Delta \nu_B}{2 \pi n f_m \Delta f}, \quad (2)$$

where $\Delta \nu_B$ is the Brillouin linewidth and $\Delta f$ is the modulation amplitude of the laser frequency. As the measurement range and the spatial resolution are in the trade-off relation, their ratio is often used as a system evaluation parameter, which is given by

$$N_{\text{BOCDR}} = \frac{\pi \, \Delta f}{\Delta \nu_B}. \tag{3}$$

Here, it is reported that $f_m$ higher than $\Delta \nu_B$ does not contribute to the enhancement of $\Delta z_{\text{BOCDR}}$ [22]. Besides, $\Delta f$ needs to be lower than a half of the BFS of the FUT because of the Rayleigh noise [20]. Therefore, the theoretical limitations of the spatial resolution and the range-to-resolution ratio can be expressed as

$$\Delta z_{\text{BOCDR}}^{\min} = \frac{c}{\pi \, n \, BFS}, \tag{4}$$

$$N_{\text{BOCDR}}^{\max} = \frac{\pi \, BFS}{2 \, \Delta \nu_B}. \tag{5}$$

If a silica SMF ($n$ = 1.46, BFS = 10.8 GHz, $\Delta \nu_B$ = 30 MHz) is used, the theoretically highest resolution and ratio are calculated to be approximately 6.0 mm and 565, respectively. The ratio can be improved by employing special schemes based on temporal gating [24] and/or dual-frequency modulation [25].

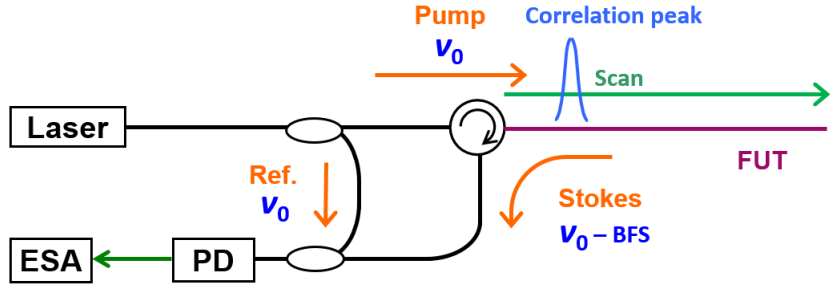

Fig. 1. Conceptual setup of basic BOCDR. ESA, electrical spectrum analyzer; FUT, fiber under test; PD photo diode.

Next, a schematic of a conceptual S-BOCDR setup is shown in Fig. 2. The laser output is directly launched into an FUT, and the reflected light is guided to a PD followed by an ESA. The reflected light consists not only of the Brillouin Stokes light but also of the Fresnel-reflected light generated at the open end of the FUT, which is exploited as the reference light. When the laser frequency is sinusoidally modulated, correlation peaks are synthesized along the FUT. The 0th correlation peak is generally fixed at a zero-optical-path-difference point, which corresponds to the open end in this case. To suppress the influence of the 0th peak, some loss is artificially applied near the open end. By sweeping the modulation frequency upward, the 1st correlation peak gradually approaches the 0th peak. As long as only the 1st peak (except the 0th peak) is in the FUT, a distributed measurement can be correctly performed. When the 1st peak has come to the middle point of the FUT, the 2nd peak starts to appear in the FUT, resulting in the erroneous measurement. Thus, the measurement range of S-BOCDR is limited to

$$d_{\text{S-BOCDR}} = \frac{L}{2}, \tag{6}$$

where $L$ is the FUT length. The spatial resolution is given, as a function of the sensing position $l$ in the FUT (see Fig. 3 for the definition of $l$), by

$$\Delta z_{\text{S-BOCDR}} = \frac{\Delta \nu_B \, (L - l)}{\pi \, \Delta f}. \tag{7}$$

Although the resolution drastically changes depending on the sensing position, we define the range-to-resolution ratio of S-BOCDR as that calculated using the lowest resolution ($l$ = 0). The ratio is then given by

$$N_{\text{S-BOCDR}} = \frac{\pi \, \Delta f}{2 \, \Delta \nu_B}. \tag{8}$$

Consequently, the theoretical limitations of the resolution and the ratio are given by

$$\Delta z^{\min}_{\text{S-BOCDR}} = \frac{2\,\Delta v_B\,(L-l)}{\pi\,BFS}, \qquad (9)$$

$$N^{\max}_{\text{S-BOCDR}} = \frac{\pi\,BFS}{4\,\Delta v_B}. \qquad (10)$$

As the 1st correlation peak is used in S-BOCDR, the ratio cannot be improved in principle by the dual-modulation technique [25]. When a 100-m-long silica SMF (BFS = 10.8 GHz, $\Delta v_B$ = 30 MHz) is used, the highest resolution is ~180 mm at $l$ = 0 m and ~90 mm at $l$ = 50 m. Note that these theoretically highest performance has been derived without taking into consideration the degradation of signal-to-noise ratio (SNR), and that it is thus difficult to achieve the highest performance experimentally.

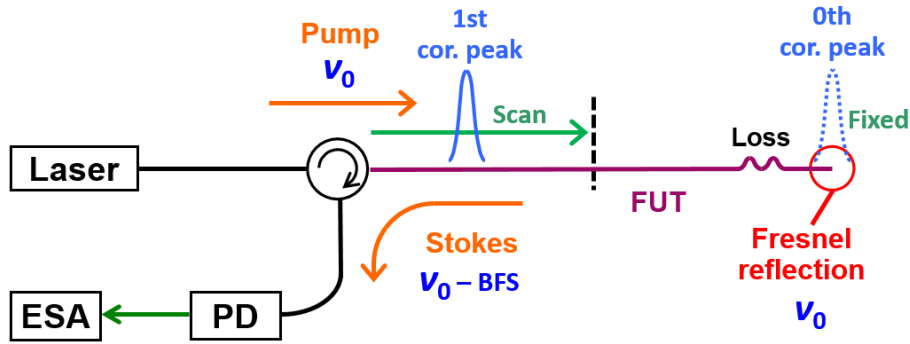

Fig. 2. Conceptual setup of S-BOCDR.

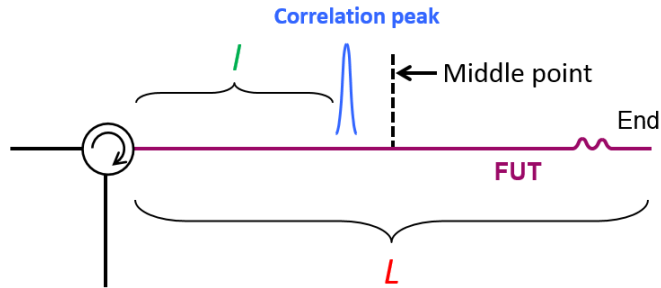

Fig. 3. Definitions of $L$ and $l$.

## 3. Experiments

The S-BOCDR setup used in the experiment is depicted in Fig. 4. The pump light was amplified up to ~30 dBm with an erbium-doped fiber amplifier (EDFA), and a 10-dB bending loss (experimentally optimized) was applied to the section near the open end. Considering that the Fresnel reflectivity at the open end which is cut vertically to the fiber axis is ~4% (= –14 dB), the reference power at the PD was roughly calculated to be –4 dBm. The sinusoidal modulation of the laser frequency was performed by directly modulating the laser driving current.

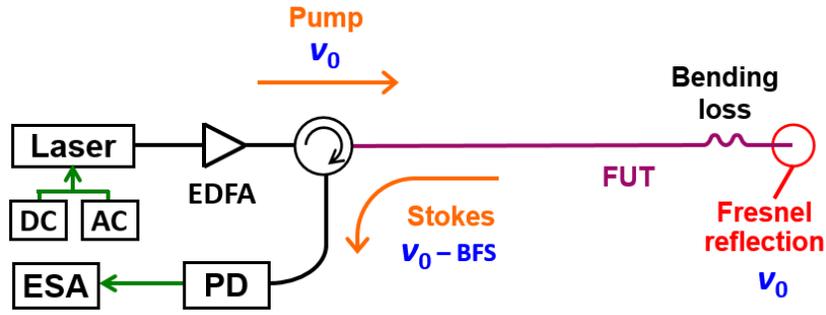

Fig. 4. Experimental setup of S-BOCDR.

We employed a silica SMF as an FUT, which had a numerical aperture of 0.13, a core refractive index of ~1.46, a core diameter of 9 μm, a cladding diameter of 125 μm, and a propagation loss of ~0.5 dB/km at 1.55 μm. The FUT length was 8.2 m, leading to the measurement range of 4.1 m. In this paper, to clearly show a distributed measurability with a high SNR, we swept the modulation frequency $f_m$ from 12.370 MHz to 24.445 MHz and fixed the modulation amplitude $\Delta f$ at 0.82 GHz, corresponding to the spatial resolution of ~95 mm at $l$ = 0 m and ~48 mm at $l$ = 4.1 m. Using two translation stages, different strains of < 0.6% were applied to a 0.6-m-long section of the FUT (see Fig. 5 for the detailed FUT structure). The sampling rate at a single sensing point was 3.3 Hz, limited by the data acquisition from the ESA. The number of the sensing points was set to 41, and consequently, the measurement time was ~12 s. The operating temperature was kept at 27 °C.

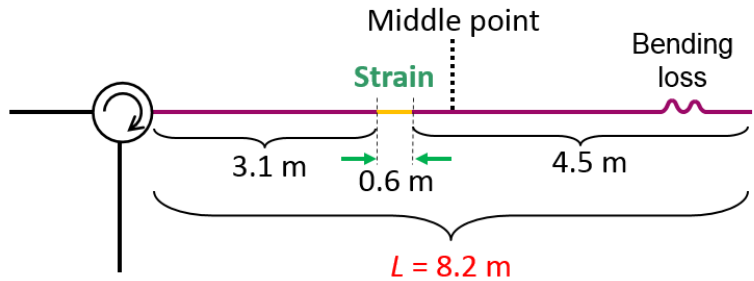

Fig. 5. Structure of FUT.

The measured BGS distribution with a 0.56-% strain applied is shown in Fig. 6. A BFS upshift was observed at the strain-applied section with a high SNR. The measured BFS distributions with strains of 0.00, 0.27, 0.40, and 0.56% applied (Fig. 7) shows that the 0.6-m-long strain-applied section was successfully detected. As shown in the inset of Fig. 7, the BFS shifted to higher frequencies with increasing strain with a proportionality constant of 491 MHz/% (calculated using the BFS values at the middle point of the strain-applied section). This value is in good agreement with the previously reported value [31]. The BFS changed even along the strain-free sections by about ±2.1 MHz (standard deviation), which indicates that the strain and temperature measurement errors are ~±0.004% and ~±1.8 °C, respectively. Thus, a distributed measurability based on S-BOCDR was demonstrated.

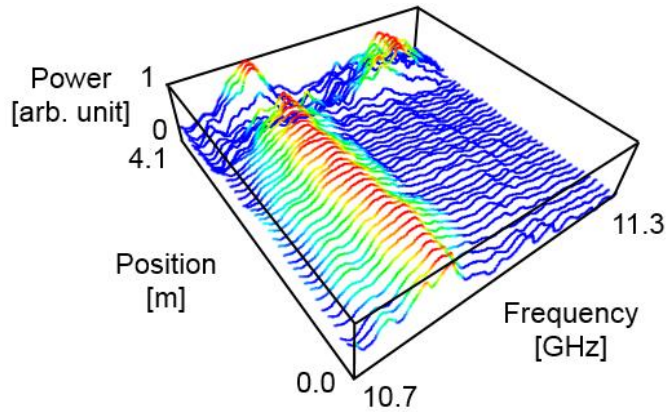

Fig. 6. Measured BGS distribution when a strain of 0.56% was applied.

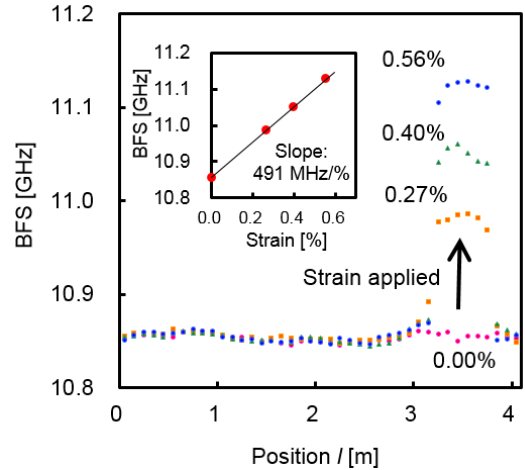

Fig. 7. Measured BFS distribution when strains of 0.00, 0.27, 0.40, and 0.56% were applied. The inset shows the BFS dependence on applied strain.

## 4. Conclusion

We developed a simple configuration of BOCDR — S-BOCDR — where the Fresnel-reflected light from an open end of a sensing fiber serves as reference light and no additional reference path is required. Frist, the theoretical limitations of spatial resolution, measurement range, and their ratio were clarified. Then, a distributed strain measurement with a spatial resolution of < 100 mm (varied depending on the sensing position) and a measurement range of 4.1 m (half the FUT length) was demonstrated with a high SNR. Here, we have to admit that S-BOCDR has the following drawbacks: (1) an FUT with a length of double the measurement range needs to be employed, (2) the range-to-resolution ratio is reduced by half, (3) if the sensing fiber is cut during the operation, the measurement cannot be continued, (4) the polarization state cannot be fully optimized using standard polarization controllers (which does not matter if polarization scrambling [33] is employed to stabilize the measurement), and (5) the system simplification does not necessarily lead to a significant reduction in cost. However, it is of considerable physical interest that a high-resolution distributed measurement can be performed by such a simple setup. Moreover, the successful demonstration of the S-BOCDR operation implies that this principle could in turn be a possible noise source for standard BOCDR systems, providing a useful guideline for improving their SNR. Thus, we believe that the development of S-BOCDR is an important technological step toward the implementation of higher-performance distributed sensors based on BOCDR.